\documentclass[doublespacing,reviewcopy]{elsart}
\usepackage{ifpdf}
\usepackage{graphicx,amssymb,lineno}
\ifpdf
\usepackage[%
  pdftitle={Instructions for use of the document class
    elsart},%
  pdfauthor={Simon Pepping},%
  pdfsubject={The preprint document class elsart},%
  pdfkeywords={instructions for use, elsart, document class},%
  pdfstartview=FitH,%
  bookmarks=true,%
  bookmarksopen=true,%
  breaklinks=true,%
  colorlinks=true,%
  linkcolor=blue,anchorcolor=blue,%
  citecolor=blue,filecolor=blue,%
  menucolor=blue,pagecolor=blue,%
  urlcolor=blue]{hyperref}
\else

\usepackage[%
  breaklinks=true,%
  colorlinks=true,%
  linkcolor=blue,anchorcolor=blue,%
  citecolor=blue,filecolor=blue,%
  menucolor=blue,pagecolor=blue,%
  urlcolor=blue]{hyperref}
\fi

\makeatletter
\def\elsartstyle{%
    \def\normalsize{\@setfontsize\normalsize\@xiipt{14.5}}
    \def\small{\@setfontsize\small\@xipt{13.6}}
    \let\footnotesize=\small
    \def\large{\@setfontsize\large\@xivpt{18}}
    \def\Large{\@setfontsize\Large\@xviipt{22}}
    \skip\@mpfootins = 18\p@ \@plus 2\p@
    \normalsize
} \@ifundefined{square}{}{\let\Box\square} \makeatother

\begin{document}

\begin{frontmatter}

\title{The inherent fluctuations of the pulsed atom laser in $F$=2 manifold of
$^{87}Rb$ atoms}

\author{Lin Xia, Fan Yang, Xiaoji Zhou and}
\author{Xuzong Chen\corauthref{cor}},
\corauth[cor]{Corresponding author.}
\address{School of Electronics
Engineering \& Computer Science, Peking University, Beijing 100871,
the People's Republic of China}

\ead{xuzongchen@pku.edu.cn, Phone: +86-10-6275-1778, Fax:
+86-10-6275-3208.}

\begin{abstract}
We have observed the intensity fluctuations of the $F$=2 $^{87}Rb$
atom laser at low output coupling rate. Theoretically, we find that
the atom loss of the condensate due to the output of atom laser
leads to fluctuations of the laser pulses, which is inherent in all
state changing out-coupling such as rf and Raman. Another reason
leading to large fluctuations is the interference of output pulses.
\end{abstract}

\end{frontmatter}

\section{Introduction}
\setlength{\parindent}{16pt}An atom laser, which is a bright,
directed and coherent beam of atoms, is quite similar to an optical
laser. As the development of optical lasers has revolutionized the
field of light optics, atom lasers have the potential to
revolutionize the field of atom optics. One of the applications of
the atom laser is precision measurement due to the potential high
sensitivity. For example, the sensitivity may be orders higher in a
matter wave gyroscope than that in a photon gyroscope \cite{ref1}.
In addition, pulsed atom lasers have been used as the mater wave
interferometer \cite{refa1,refa2} and the detector to study the
characters of Bose-Einstein condensates (BEC)
\cite{refa3,refa4,refa5}. Because of these applications, to study
the fluctuations of the pulsed atom laser becomes important, which
is the first step to get rid of or avoid the fluctuations. Recent
studies do show that the fluctuations are large under certain
conditions. The fluctuations due to the coupling back of output
beams in the rf coupled atom laser system are reported \cite{ref2}.
The strong strength of output coupling leads to fluctuations even
'shut down' of the laser \cite{ref3}. The differences between
two-state and multistate atom laser are studied, and the sloshing of
the middle state atoms induce noise of continuously out-coupled atom
laser \cite{ref4}.

Recently, many works
\cite{refad1,refad2,refad3,refad4,refad5,refad6} on spinor BEC
appeared. The complex structure of internal states is a condition
that enable the spinor BEC system to reveal rich dynamics.
Similarly, the $F=2$ system, a five-state system, is expected to
provide interesting and rich dynamics during the output of the atom
laser.

In this paper, we study the various fluctuations at low coupling
rate where the back coupling induced fluctuations \cite{ref2} are
avoided. Compared with the previous works, we find the following
interesting results: (1) under the condition that the back-coupling
induced fluctuations can be neglected, the resolvable picture in the
experiment demonstrates large fluctuations. (2) with the
Gross-Pitaevskii (GP) equations, we demonstrate two new reasons
leading to fluctuations: the atom loss of the condensate due to the
output of atom laser and the interference of the output atom clouds.
(3) our results show that the fluctuations of the atom laser in the
$F$=2 manifold of $^{87}Rb$ atoms do not decrease with the coupling
strength, which is different from the continuously coupled atom
laser.

\section{The experimental results}
The experimental setup is the same as that described in our previous
work \cite{ref5}. In brief, we get samples of condensates in a
compact low power quadrupole-Ioffe-configuration (QUIC) trap with
trapping frequency $\omega_r=2\pi\times225$ Hz radially and
$\omega_z=2\pi\times20$ Hz axially. The current is 20.7 A in
quadrupole coils and 20.5 A in Ioffe coil. Typically, a $^{87}Rb$
condensate with $5\times10^4$ atoms in $|F=2$, $m_{F}=2\rangle$
state is formed after the evaporative cooling. Then the rf pulses
are switched on to couple the atoms. The images are taken right
after switching off the magnetic trap.

Fig. 1 shows the image of atom laser after 4 rf pulses. The time
separations are $\Delta t_1=1.9 $ ms between the first and second
pulse, $\Delta t_2=2.2 $ ms between the second and the third pulse,
$\Delta t_3=2.9 $ ms between the third pulse and the fourth pulse
and $\Delta t_4=5.0 $ ms between the fourth pulse and the imaging.
The back-coupling induced noise \cite{ref2} can be neglected because
of the large enough time separations. The rf pulse duration is 24 $
\mu s$, and the coupling strength is $\Omega =19000 $ Hz. The
frequency of the rf is set to be resonant with the center of the
condensate. In Fig. 1, the first out-coupled pulse is a single could
of atoms as we expect. However, the second one is composed of two
clouds. The third and the fourth one look more complex. The
fluctuations appear because the $m_F=1$ state (trapped state) is
populated after the first rf pulse. Then the following atom laser
pulses come from both the $m_F=2$ and $m_F=1$ state. This has been
predicted in \cite{ref2} theoretically. However, it has not been
observed experimentally.

\section{Analyses using the Gross-Pitaevskii equations}
To investigate the forming process of atom laser beams, the
one-dimensional GP equations \cite{ref3, ref6} are used to study the
dynamics of the coupling process, which are given by
\begin{eqnarray*}
i\dot{\psi_2}&=&(-\frac{1}{2}\partial ^2/{\partial
y}^2+y^2+2\Delta-G+U(\sum_{i=-2}^2{\vert\psi_i\vert}^2))\psi_2+2\Omega\psi_1
\\ i\dot{\psi_1}&=&(-\frac{1}{2}\partial ^2/{\partial
y}^2+\frac{1}{2}y^2+\Delta-G+U(\sum_{i=-2}^2{\vert\psi_i\vert}^2))\psi_1+2\Omega\psi_2+\sqrt{6}\Omega\psi_0
\\ i\dot{\psi_0}&=&(-\frac{1}{2}\partial ^2/{\partial
y}^2-G+U(\sum_{i=-2}^2{\vert\psi_i\vert}^2))\psi_0+\sqrt{6}\Omega\psi_1+\sqrt{6}\Omega\psi_{-1}
\\ i\dot{\psi_{-1}}&=&(-\frac{1}{2}\partial ^2/{\partial
y}^2-\frac{1}{2}y^2-\Delta-G+U(\sum_{i=-2}^2{\vert\psi_i\vert}^2))\psi_{-1}+\sqrt{6}\Omega\psi_0+2\Omega\psi_{-2}
\\ i\dot{\psi_{-2}}&=&(-\frac{1}{2}\partial ^2/{\partial
y}^2-y^2-2\Delta-G+U(\sum_{i=-2}^2{\vert\psi_i\vert}^2))\psi_{-2}+2\Omega\psi_{-1}.
\end {eqnarray*}

\setlength{\parindent}{-2pt} Here
$G=g(M/\hbar\omega{_{1y}^3})^{1/2}$ is the gravity, where
$\omega_{1y}$ is the radial trapping frequency of the $m_F$=1 state.
$U={g_{1d}M^{{1}/{2}}}/({\hbar^{{3}/{2}}\omega{_{1y}^{{1}/{2}}}})$
is the interaction coefficient, where $g_{1d}$ is the effective
interaction strength. $\Delta$ is the detuning of rf field from the
resonance of atoms at the center of the magnetic trap. The time,
spatial coordinates, wave function and detuning are measured in the
units of $\omega_{1y}^{-1}$, $({{\hbar}/{M\omega{_{1y}}}})^{1/2}$,
$({{\hbar}/{M\omega{_{1y}}}})^{-1/4}$ and $\omega_{1y}$,
respectively. The large velocities of atoms due to the falling down
in gravitational potential make the numerical descriptions complex
\cite{ref4}. The small de Broglie wavelengths due to the large
velocities require very fine temporal and spatial grids to ensure
the validity of numerical simulation. We use the 1D GP equations to
simplify the numerics and the time-splitting sine-spectral (TSSP)
method \cite{ref7} preformed by FORTRAN to speed up the simulation.
Fig. 2 shows the experimental data of the 1D density and the
numerical fitting. Picture (a),(b),(c) and (d) correspond to the
first, second, third and fourth pulse at $t=12$ ms, respectively. In
fact, it appears that there is more information in the numerical
simulation results since the resolution of absorption image is
limited to several $\mu m$ typically. We use the GP equations to
study the fluctuations in the following, which removes the
limitation of the image resolution.

The forming process of the atom beams without the back-coupling is
shown in Fig. 3 by numerically solving the 1D GP equations. Although
the process with back-coupling has been discussed in \cite{ref2}, we
provide the analysis to better understand the following discussions.
The parameters are the same as that in experiments. In the first
column, we can see the emergence of $m_F$=1 component after the
first rf pulse. Because the equilibrium position of the $m_F$=1
state is a bit lower than that of the $m_F$=2 state, the new $m_F$=1
atoms oscillate around its equilibrium position (the $m_F$=2 state
and $m_F$=1 state atoms separate in space). When the second rf pulse
is switched on, both the $m_F$=2 state and the $m_F$=1 state are
coupled into the $m_F$=0 state, which induces that the second atom
laser pulse is composed of two clouds of atoms. Note that there are
two clouds atoms in both $m_F$=2 component and $m_F$=1 component
after the second pulse. This induces that the third atom laser pulse
is composed of 4 pulses in column 4 (two of the clouds overlap). So
the nth output atom laser pulse is composed of $2^{n-1}$ atom clouds
in principle, which means that the fluctuations increase very fast
with coupling times. In this process, the existence of the $m_F$=1
state creates the fluctuations, and the population transfer between
the $m_F$=2 state and the $m_F$=1 state aggravates the fluctuations
quickly.



%

Using the GP equations, the fluctuations with different coupling
strength are shown in Fig. 4. The fluctuations keep large as the
decreasing of the coupling strength. The shape of the density
distribution trends to be fixed (Fig. 4 (c) and Fig. 4 (d)) at very
low coupling strength. This is different from the continuously
coupled atom laser. For continuously coupled atom laser, experiments
demonstrate that quite stable flux can be formed if the coupling
strength is low \cite{ref3,ref8}, and numerical simulations also
show that the fluctuations decrease with the coupling strength
\cite{ref4}. This can be explained by the difference of the spatial
resonant width of the output coupling. A 20 $ \mu s$ rf pulse has a
frequency width of 100 kHz corresponding to 20 $\mu m$ spatial width
of resonance. The $m_F=1$ component, which originated from the
$m_F=2$ state, can not escape from this large resonant region before
next rf pulse. So the $m_F=0$ component, which is originated from
the mF=1 component, becomes the fluctuations of the output laser
pulse. The fluctuations occur whether the coupling strength is high
or low. In continuously coupling case, a 10 ms rf pulse has a
frequency width of 200 Hz corresponding to 0.2 $\mu m$ spatial width
of resonance. The $m_F=1$ component, which is originated from the
$m_F=2$ state, leaves the resonant region quickly before inducing
large fluctuations. In addition, the 1D GP equations appear to be a
useful tool to study the dynamics of atom lasers. The atoms numbers
in Fig. 4 (b), (c) and (d) are estimated to be at least one order
less than that in Fig. 4 (a) so it is hard to study experimentally.

We find that the atom loss of the condensate due to the output of
atom laser leads to fluctuations. Fig. 5 (a) shows the out-coupled
atom lasers with different time separation $\Delta t_1$ between the
first and second pulse. Plotted are the atom laser pulses generated
from the condensate ($m_F$=2 component) right after the second
pulse. From the top to bottom the rows are the time separation
$\Delta t_1=$1.90 ms, 2.49 ms, 3.08 ms, 3.67 ms and 4.25 ms
respectively. From the left to right the columns are the coupling
strength $\Omega=$19000 Hz, 12000 Hz and 2000 Hz. The two rf pulses
in each process are with the same strength. The shape of the
generated atom laser changes periodically with the coupling time at
high coupling strength. The amplitude of the fluctuations decreases
with the coupling strength. The explanation of this phenomenon can
be found in Fig. 5 (b), where the evolution of the 1D density of the
condensate as time after the first pulse is shown. The rf pulse is
fired at $t$=0. The coupling strength is $\Omega=$19000 Hz, 12000 Hz
and 2000 Hz from left to right, which is the same as that in Fig. 5
(a). When a large quantity of atoms are out-coupled from the
condensate, the mean-field interaction potential between an atom and
the condensate changes. The condensate does not stay in the original
equilibrium any more. The 1D density shape of the condensate changes
periodically with time. The density shape of the extracted atom
laser by the following rf pulse also changes periodically with time
because the generated pulse retains the density spread of the
condensate in space. When the atom loss is quite few (the right one
of Fig. 5 (b)) so that the change of the interaction potential is
quite small, the fluctuations of the atom laser can be neglected.
The condensate is not continuously repumped in recent atom laser
experiments so this kind of fluctuation is expected to be inherent
in all state changing out-coupling, e.g., rf and Raman.

It appears that the interference of out-coupled atom clouds is
another new reason leading to large fluctuations. The atom laser
pulse is composed of several atom clouds except the first pulse.
Interference occurs when different clouds overlap in space. Fig. 6
shows the process that interference appears by solving the 1D GP
equations. The parameters are the same with the experiment. From
left to right in Fig. 6 (a), the first atom cloud is out-coupled
from $m_F$=2 state, and the second is from $m_F$=1 state. There is a
relative velocity between them. As they overlap, stripes appear in
Fig. 6 (b). Then they sufficiently overlap later in Fig. 6 (c).
These interference stripes are with high frequency in space, which
make them hard to be detected by absorption image, e.g., in the Fig.
2 (d). The fluctuations trend to be larger in later laser pulse in
Fig. 2 as more clouds in one pulse have more chances to overlap.

\section{Conclusion}
In conclusion, we demonstrate that the fluctuations of pulsed atom
laser in the $F$=2 manifold are still large on condition that the
back coupling induced fluctuations are avoided. The obvious
fluctuations are experimentally observed. By using the 1D GP
equations, we find two new reasons leading to fluctuations: (1) the
atom loss of the condensate leads to fluctuations of the atom laser;
(2) the interference of out-coupled atom clouds is another reason
which induces large fluctuations.


\section*{Acknowledgement}
The authors thank Professor W. M. Liu at Institute of Physics of
Chinese Academy of Sciences for comments and helpful discussions.
This work was supported by the state Key Development Program for
Basic Research of China (No. 2005CB724503, 2006CB921401 and
2006CB921402) and NSFC (No. 60490280 and 10574005).

\bibliographystyle{elsart-num}


\newpage
\textbf{Figure Captions} \\
Fig. 1.  Image of the pulsed atom laser produced in the F=2
manifold.
\\
\\
Fig. 2.  The experimental data of the 1D density and the numerical
fitting at $t=12ms$.
\\
\\
Fig. 3.  The out-coupling process of the pulsed atom laser in $F$=2
manifold.
\\
\\
Fig. 4.  The fluctuations of pulsed atom laser with different
coupling strength.
\\
\\
Fig. 5.  Fluctuations of the atom laser induced by the atom loss of
the condensate.
\\
\\
Fig. 6.  Fluctuations induced by the interference of out-coupled
atom clouds.

\newpage
\begin{figure}[h]
\begin{center}
\includegraphics[width=3.3cm]{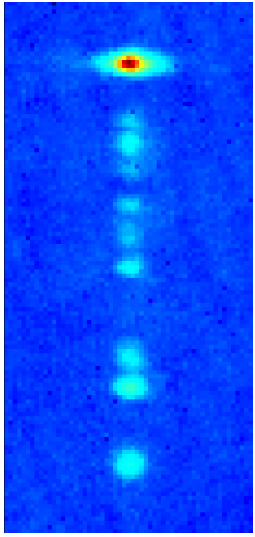}
\end{center}
\caption{\label{label}Image of the pulsed atom laser produced in the
F=2 manifold: The image shows the atom laser after 4 rf pulses. The
time separation is $\Delta t_1=1.9$ ms between the first and second
pulse, $\Delta t_2=2.2 $ ms between the second and the third pulse,
$\Delta t_3=2.9 $ ms between the third pulse and the fourth pulse
and $\Delta t_4=5.0 $ ms between the fourth pulse and the imaging.
Except the first pulse, the output atom laser pulses are with
fluctuations.}
\end{figure}

\begin{figure}[h]
\begin{center}
\includegraphics[width=8cm]{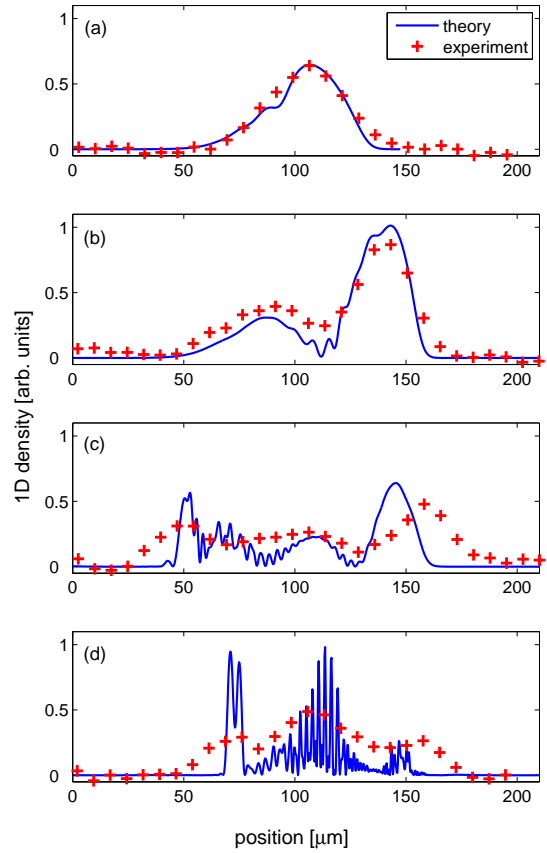}
\end{center}
\caption{\label{label} The experimental data of the 1D density and
the numerical fitting at $t=12ms$: Picture (a), (b), (c) and (d)
correspond to the first, second, third and fourth pulse,
respectively. }
\end{figure}

\begin{figure*}
\includegraphics[width=16.5cm]{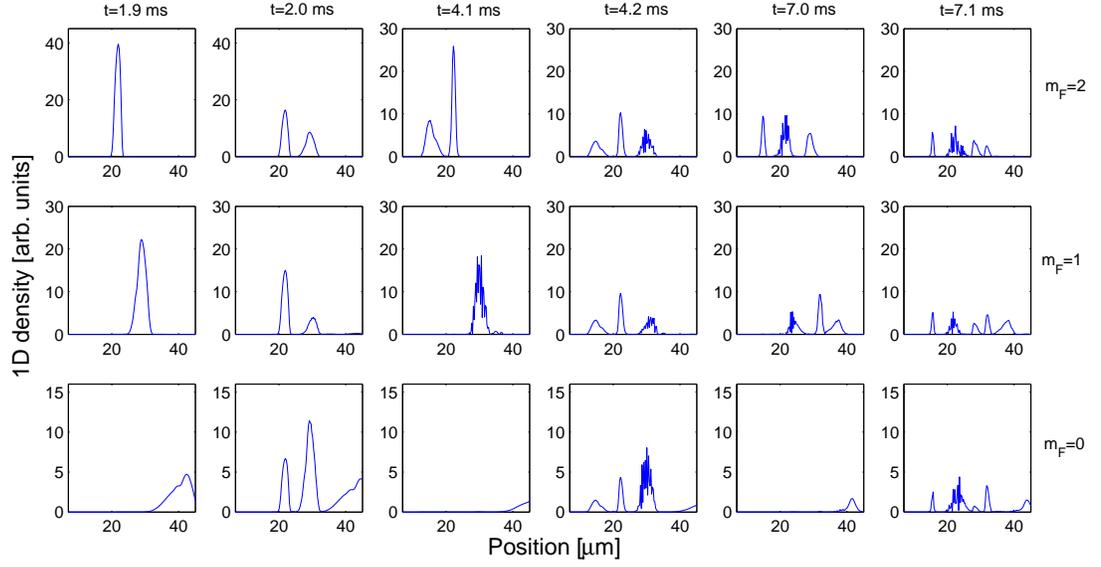}
\caption{\label{fig:wide}The out-coupling process of the pulsed atom
laser in $F$=2 manifold: From the left to right the columns are
right before the second pulse (t=1.9 ms), after the second pulse
(t=2.0 ms), right before the third pulse (t=4.1 ms), after the third
pulse (t=4.2 ms), right before the fourth pulse (t=7.0 ms) and after
the fourth pulse. From the top to bottom the rows are $m_F$=2,
$m_F$=1 and $m_F=0$ state. The gravitation is along the x axis.}
\end{figure*}

\begin{figure}[h]
\begin{center}
\includegraphics[width=6.5cm]{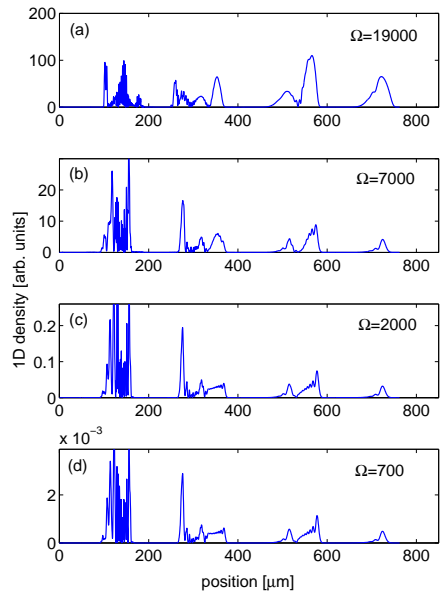}
\end{center}
\caption{\label{label} The fluctuations of pulsed atom laser with
different coupling strength: Picture (a), (b), (c) and (d)
correspond to the coupling strength $\Omega$=19000 Hz, 7000 Hz, 2000
Hz and 700 Hz, respectively. The other parameters are the same as
the experiment.}
\end{figure}

\begin{figure}[h]
\begin{center}
\includegraphics[width=9.0cm]{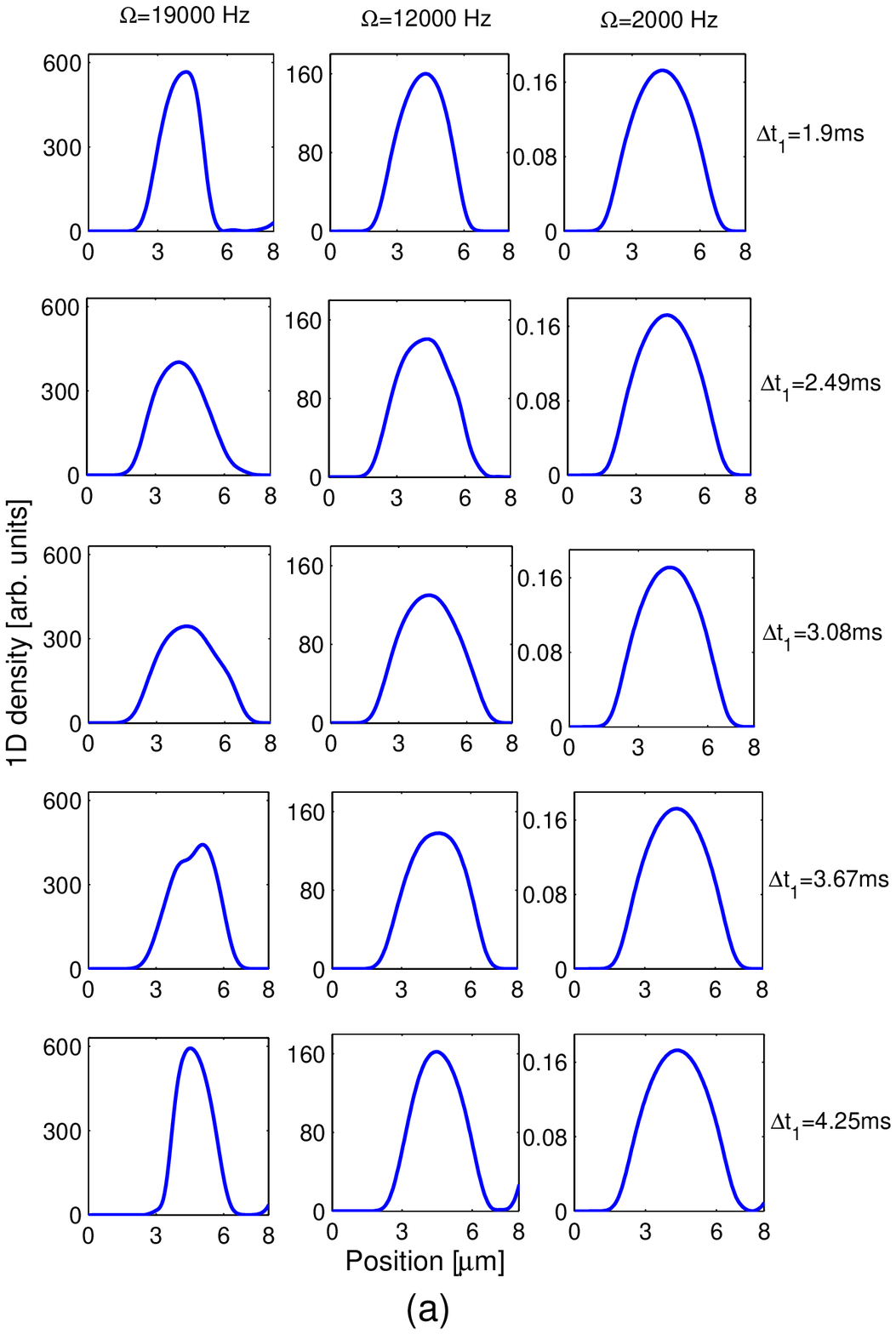}
 \includegraphics[width=9.1cm]{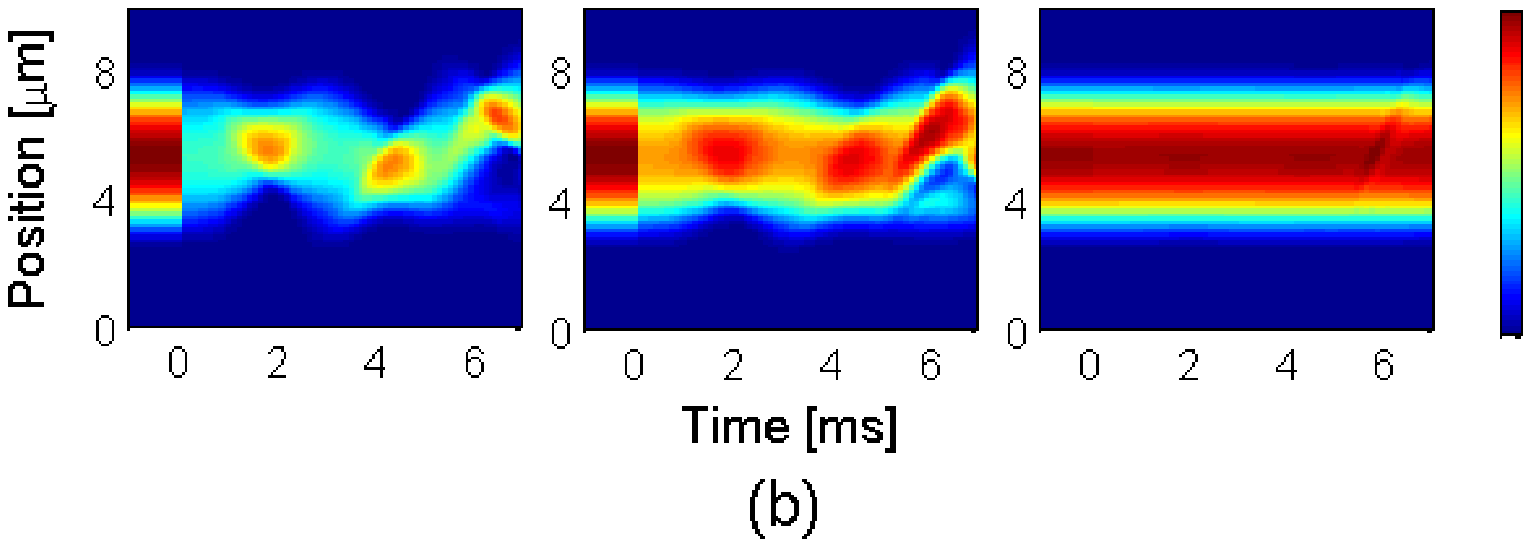}
\end{center}
\caption{\label{label}Fluctuations of the atom laser induced by the
atom loss of the condensate. (a) The 1D density of out-coupled atom
lasers with different time separation $\Delta t_1$ between the first
and second pulse. The atom laser pulse is generated from the
condensate right after the second pulse. From the top to bottom the
rows are the time separation $\Delta t_1=$1.90 ms, 2.49 ms, 3.08 ms,
3.67 ms and 4.25 ms, respectively. From the left to right the
columns are the coupling strength $\Omega=$19000 Hz, 12000 Hz and
2000 Hz. (b) The evolution of the 1D density of the condensate with
time after the first pulse. The rf pulse is fired at $t$=0. The
coupling strength is $\Omega=$19000 Hz, 12000 Hz and 2000 Hz from
left to right.}
\end{figure}

\begin{figure}[H]
\begin{center}
\includegraphics[width=7.0cm]{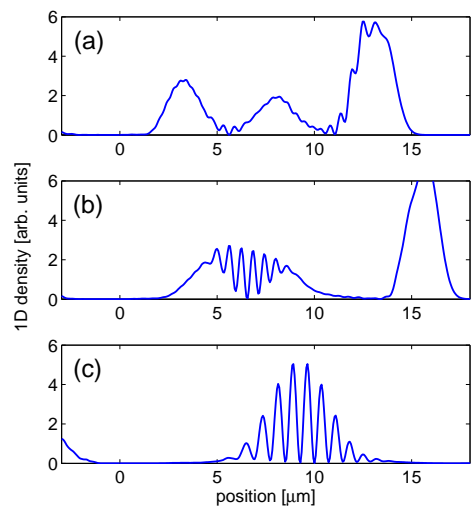}
\end{center}
\caption{\label{label} The interference induced fluctuations of
out-coupled atom clouds: Plotted is the 1D density of the third atom
laser pulse. The parameters are the same as the experiments. The
picture Fig.6 (a), (b) and (c) represent t=4.5 ms, 4.8 ms and 5.3
ms, respectively.}
\end{figure}

\end{document}